\documentclass[12pt,letterpaper]{iopart}
\usepackage{iopams,epsfig,bm}

\def\bp{{\mathbf p}}
\def\br{{\mathbf r}}
\def\bk{{\mathbf k}}
\def\bu{{\mathbf u}}
\def\rG{{\mathrm G}}
\def\mA{{\mathcal{A}}}
\def\expp{{\mathrm e\,}}

\begin{document}

\paper[Relativistic and retardation effects in the two--photon ionization
of H-like ions]
{Relativistic and retardation effects in the two--photon ionization
 of hydrogen--like ions}

\author{Peter Koval\,%
\footnote[1]{To whom correspondence should be addressed
(kovalp@physik.uni-kassel.de)},
Stephan Fritzsche and Andrey Surzhykov
}

\address{Fachbereich Physik, Universit\"at Kassel, 
         Heinrich--Plett Str. 40, D--34132 Kassel, Germany}

\begin{abstract}
The non--resonant two--photon ionization of hydrogen--like ions
is studied in second--order perturbation theory, based on Dirac's equation.
To carry out the summation over the complete Coulomb spectrum, a Green's 
function approach has been applied to the computation of the ionization cross
sections. Exact second--order relativistic cross sections are compared 
with data as obtained from a relativistic long--wavelength approximation 
as well as from the scaling of nonrelativistic results. For high--$Z$ ions, the
relativistic wave function contraction may lower the two--photon ionization 
cross sections by a factor of two or more, while retardation effects appear
less pronounced but still give rise to non-negligible contributions.
\end{abstract}

\section{Introduction}

The multi--photon ionization of atoms has been widely studied during the last 
few decades. While, however, most previous atomic experiments focused on the 
multi--photon ionization of the valence--shell electrons of the alkaline metal
and group I elements (Jaouen \etal 1984, Antoine \etal 1996), 
theoretical investigations---instead---often dealt with the excitation and 
ionization of low--Z, hydrogen-- and helium--like ions, owing to their 
simplicity (Karule \etal 1985, Maquet \etal 1998). 
With the recent progress in the development 
and set--up of coherent light sources in the EUV and x--ray domain, such as 
the various free--electron lasers, it now become much more likely that two-- 
and multi--photon processes can be observed also for the  inner--shell 
electrons of medium and heavy elements in the near future (Kornberg \etal 2002).
Since, generally, a relativistic theory is needed to describe such elements,
a primary interest in studying multi--photon processes may concern first the
importance of relativistic effects along the hydrogen isoelectronic sequence.
In the past, similar investigations have been carried out only for the decay 
of the $2s_{1/2}$ metastable level (Santos \etal 2001) as well as for the 
two--photon excitation from the $1s$ ground states of hydrogen--like ions
(Szymanowski \etal 1997). To the best of our knowledge, however, no attempt 
has been made so far to explore two-- and multi--photon processes for medium 
and high--$Z$ ions by means of a relativistic theory.

In this paper, we consider the two--photon ionization of hydrogen--like 
ions in second--order perturbation theory, based on Dirac's equation. 
To obtain the total ionization cross sections, a Green's function 
approach is applied in section \ref{theory} to perform
the summation over the  \textit{complete hydrogen spectrum} appropriately. 
Using such an approach, cross sections for the two--photon ionization of
the $1s$ ground state of hydrogen--like ions are calculated for nuclear
charges in the range $Z\,=\,1,\,\ldots,\, 100$ in order to explore both, 
the relativistic contraction of the wave functions as well as those effects
which arise from the higher multipoles in the decomposition of the radiation
field, i.~e.\ the so--called \textit{retardation effects}. 
Section \ref{results}, later, provides a comparison of the cross sections from
the relativistic theory (as obtained in two different approximations)
as well as from the scaling of non-relativistic results. Finally,
a few conclusions are given in section \ref{conclusion}.

\section{Two-photon ionization cross section. Perturbative treatment}
\label{theory}

In second--order perturbation theory, the two--photon ionization cross 
section $\sigma_2$ is given by (Laplanche \etal 1976)

\begin{eqnarray}
   \label{crosssection}   
   \sigma_2 & = & \frac{8\pi^3 \alpha^2}{ E_{\gamma}^{\,2}} \ 
   \left\vert \, 
   \sum\mkern-22mu\int_{\nu} \
   \frac{\langle \psi_{f} \ | \ \bm{p}\cdot{\bf u}_{\lambda_2} 
   e^{i\,\bk_2\cdot\br} \ | 
   \ \psi_{\nu} \rangle
   \langle \psi_{\nu} \ | \ \bm{p}\cdot {\bf u}_{\lambda_1} 
   e^{i\,\bk_1\cdot\br} \ | \ 
   \psi_i \rangle }{ E_{\nu} - E_i - E_{\gamma}}
    \, \right\vert^2 , 
\end{eqnarray}
where $(\psi_i,\, E_i)$, $(\psi_{\nu},\, E_{\nu})$ and $(\psi_f,\, E_f)$ 
denote the wave functions and the energies of the initial, intermediate and 
final atomic states, respectively\,\footnote{
Here and in the following, we use Hartree atomic units. Since the  
two--photon ionization cross section $\sigma_2$ has the dimension 
length$^4 \,\times\,$time, it can easily be converted also to other units
such as cm$^4 \cdot$ sec by using the \textit{conversion factor} 
1.89679 $\times$ 10$^{-50}$.}.
In this expression, as usual, the electron--photon interaction is described 
in terms of the transition operator 
$\bp\,\cdot\, \bu_{\lambda}\, e^{i\,\bk\br}$ 
which includes the momentum $\bp$ of the electron and the photon wave 
$\bu_{\lambda}\,e^{i\bk\cdot\br}$. As appropriate for
laser experiments, here and in the following we assume that both 
photons have equal wave vectors $\bk_1\,=\,\bk_2\,=\,\bk$ and equal helicities
$\lambda_1\,=\,\lambda_2\,=\,\lambda\,=\,\pm\,1$, i.~e.\ that they have the 
same circular polarization. Then, the energy 
$E_f \,=\, E_i \,+\, 2 E_{\gamma}$ of the emitted electron simply follows from
the energy conservation and is given by the energy of the initial state and 
twice the photon energy $E_{\gamma}$.

\subsection{Green's function method}

Apart from the usual integration over the spatial coordinates, 
the evaluation of the transition amplitude in in Eq.~(\ref{crosssection})
also requires a \textit{summation} over the complete spectrum of the (hydrogen)
ion. Obviously, this summation includes the sum over all discrete states 
as well as an integration over continuum. In particular
the second part, i.~e.\ the integration over the continuum, 
is rather difficult to carry out in practice since it 
implies the computation of \textit{free--free} electronic transitions. 
An alternative to carrying out the summation over the spectrum explicitly
in the transition amplitude is given by a change in the sequence of 
summation and integration from $\int\int \, d\br\,dE_{\nu}$ to 
$\int\int \, dE_{\nu}\,d\br$. Then, the summation over the complete
hydrogen spectrum can be replaced by the Coulomb Green's
function (Swainson and Drake, 1991)
\begin{eqnarray}
\rG_E(\br,\, \br')\,=\, \sum\mkern-22mu\int_{\nu}
\frac{|\psi_{\nu}(\br)\rangle\langle\psi_{\nu}(\br')|}
     {E_{\nu}\,-\,E} \, 
\label{expansion}
\end{eqnarray}
which is zero at the origin and 
tends to zero if $r \rightarrow \infty$ or $r' \rightarrow \infty$,
respectively. This particular property of Coulomb Green's function
ensures that  the second-order transition amplitudes 
in (\ref{crosssection}) can be evaluated even if the
continuum wavefunctions $\psi_f$ remains oscillating at large $r$.

Using the Green's function \ref{expansion}, the ionization 
cross section (\ref{crosssection}) can be re-written 
in the form (Maquet \etal 1998)
\begin{equation}
   \label{cs_greens}
   \sigma_2 \ = \ \frac{8\pi^3 \alpha^2}{ E_{\gamma}^{\,2}} \ 
   \left\vert \, 
   \langle \psi_{f} \ | \ \bm{p}\,\cdot\, \bu_{\lambda} e^{i\bk\cdot\br} \ 
   \rG_{E_i + E_{\gamma}}(\br, \br')
   \ \bm{p'}\,\cdot\, \bu_{\lambda} e^{i\bk\cdot\br'} \ | \ 
   \psi_i \rangle\right\vert^2,
\end{equation}
including a new \textit{double} integration 
over $\br$ and $\br'$. For hydrogen--like ions, the Coulomb--Green's 
functions $\rG_{E}(\br,\,\br')$ are known analytically, both within the 
nonrelativistic as well as the relativistic theory. Based on the 
Dirac--Hamiltonian with a hydrogen potential, 
$H_{\rm D} \,=\, c \bm{\alpha} \cdot \bm{p}\, +\,\beta mc^2 \,-\,Z/r$,
a radial--angular representation of the relativistic Coulomb--Green's function
was given earlier by Swainson and Drake (1991). In the evaluation of matrix
elements, such a representation allows for the analytic integration 
over all angles by using the techniques of Racah's algebra, while the radial
integration has---often---to be carried out numerically.

\subsection{Multipole expansion of the photon wave}

\label{multexp}

To evaluate the \textit{angular part} of the transition amplitude in expression
(\ref{cs_greens}), of course, we need first to represent the photon wave in 
terms of its electric and magnetic multipole fields (Rose 1957)
\begin{equation}
   \label{photon_expansion}
\bu_{\lambda} \expp^{ikz}= 
    \sqrt{2\pi} \ \sum_{L=1}^{\infty} \ i^L \ \sqrt{2L+1} \ 
    \left( \ \mA_{L \lambda}^{(m)} + i \lambda \mA_{L \lambda}^{(e)} \ \right),
\end{equation}
where, for the sake of simplicity, we have taken the quantization axis, i.~e.\
the  $z$--axis, along the photon momenta $\bk$. For a proper 
radial--angular representation of all Coulomb wave and Green's functions,
then, the transition amplitude can be reduced to a (finite) sum  of products 
of the type \textit{angular coefficient $\times$ radial integral\,}, 
in dependence on the number of multipoles and on further approximations 
which are made for the (coupling of the) radiation field. In our computations,
the angular coefficients were obtained algebraically, using the \textsc{Racah} 
program (Fritzsche 1997, Fritzsche \etal 2001).  For the radial integrals,
in contrast, we applied the procedures from the \textsc{Greens} 
library (Koval and Fritzsche 2003). 
Owing to the structure of the radial Green's function (matrix) this implies 
a double integration over a 2--dimensional area in 
$0 \,\le \, r \,<\,\infty$ and $0 \,\le \,r'\,\le\,\infty$, for which an adaptive 
numerical integration algorithm with a user--defined precision was
developed. This algorithm is based on the Gauss--Legendre quadrature and has
been implemented as well in the \textsc{Greens} library.

\section{Results and discussion}

\label{results}

\subsection{Relativistic Z-scaling rule}

Different approximations can be applied to investigate the two--photon
ionization of hydrogen--like ions, in dependence on the photon frequency and
the nuclear charge. In nonrelativistic quantum theory, for instance,
the total non--resonant cross section in the long--wavelength approximation 
is known to scale down like
\begin{eqnarray}
\label{nrel_scaling}
   \sigma_2(Z,\,E_{\gamma} \cdot Z^2) & = &
   \frac{1}{Z^{6}} \,\cdot\, \sigma_2(Z=1,\,E_{\gamma}) \, , 
\end{eqnarray}
i.~e.\ with the sixth power of the nuclear charge, if---at the same time---the
photon energy is scaled with $Z^{\,2}$ (Zernik 1964). This scaling rule for the
non--resonant part of the cross section applies for all photon energies 
$\mathrm{Ry}/2 \,\le\, E_{\gamma} \,<\,\mathrm{Ry}$ below of the one--photon
threshold of hydrogen $(Z =1)$, where $\mathrm{Ry} \,\simeq\, 13.6$ eV refers 
to the hydrogen ground--state energy. To display the deviations of the cross 
sections in the different relativistic approximations from the 
nonrelativistic scaling, we may re--write Eq.\ (\ref{nrel_scaling}) in the 
form
\begin{eqnarray}
\label{scaling_rel}   
   \sigma_2(Z,\, E_\gamma(Z)) & = &
   \frac{\xi(Z)}{Z^6}\cdot\sigma_2(Z=1,\, E_\gamma(Z=1)),
\end{eqnarray}
where the photon energy 
$E_{\gamma}(Z) \,\equiv\, \varepsilon \,\cdot\, |E_{1s}(Z)|/2$ now depends
on the relativistic binding energy and, thus, shows a slightly more complicated
$Z-$dependence than the nonrelativistic $\sim Z^{\,2}$ behaviour. As above, 
we may restrict ourselves to photon energies with 
$1 \,\le\, \varepsilon \,<\,2$ below of the one--photon threshold of all
hydrogen ions. With this definition of $\varepsilon$, however, the 
interpretation of the scaling rule (\ref{scaling_rel}) becomes quite simple 
as, say, a value $\varepsilon \,=\, 1.05$ obviously specifies the photon energy
so, that the total energy of the two photons together exceeds the $1s$ 
threshold by just 5\%{}; a definition which can be used also in the 
nonrelativistic case. Thus, the \textit{net} deviation between the various 
approximations is shown in the scaling factor $\xi(Z)$ which, in the 
nonrelativistic limit, is $\xi(Z) \,\equiv\, 1$.

\begin{figure}
\begin{center}
\epsfig{file=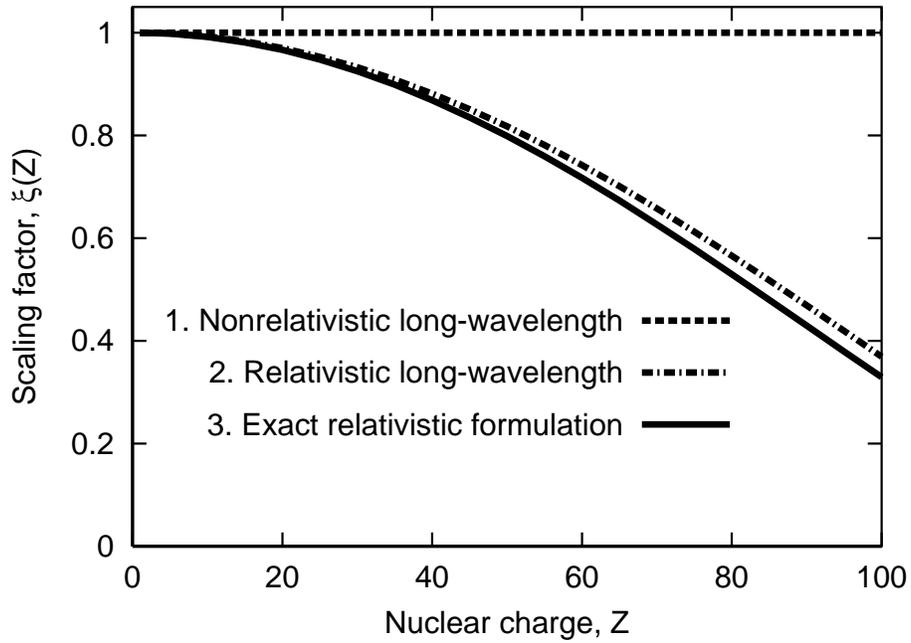, height=12cm, angle=270}
\caption{Dependence of the scaling factor $\xi(Z)$ on the nuclear charge $Z$
for $\varepsilon \,=\, 1.05$, i.~e.\ for a two--photon excess energy of 5~\%{}. 
1.~Nonrelativistic long--wavelength approximation;
2.~Relativistic long-wavelength approximation; 
3.~Exact relativistic second--order results.}
\label{f:low_energy}
\end{center}
\end{figure}

\subsection{Relativistic and retardation effects}

Figure \ref{f:low_energy} displays the scaling factor $\xi(Z)$ as function
of the nuclear charge $1 \,\le\, Z \,\le\, 100$ for $\varepsilon \,=\,1.05$,
i.~e.\ for a two--photon excess energy of 5~\%{} which is well within the 
non--resonant region. Three different approximations are shown in this figure: 
Apart from the trivial nonrelativistic factor $\xi(Z)\,=\,1$, the scaling 
factors are given for the relativistic long--wavelength  approximation
$e^{i\bk\br} \,=\,1$ (dashed--dotted line) as well as for the exact second--order 
perturbation treatment of all retardation effects (solid line). In practice, 
only the multipole fields up to $L_{\rm max} \,=\, 5$ 
are needed in (\ref{photon_expansion}) in order to obtain convergence of the
corresponding cross sections at about the 1\% level.

\begin{figure}
\begin{center}
\epsfig{file=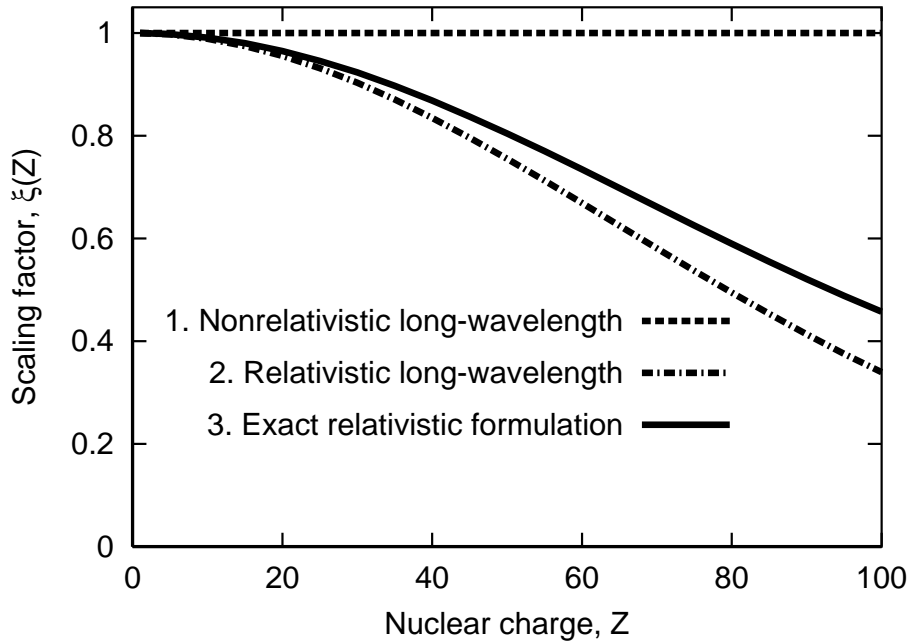, height=12cm, angle=270}
\caption{Dependence of the scaling factor $\xi(Z)$ on the nuclear charge $Z$
for $\varepsilon \,=\, 1.40$. All other notations are the same as in Figure 
\ref{f:low_energy}. }
\label{f:high_energy}
\end{center}
\end{figure}

When compared with the nonrelativistic decrease of the two--photon ionization
cross sections, owing to the $1/Z^{\,6}$ scaling of the cross sections in
Eq.\ (\ref{nrel_scaling}), a further significant reduction arises for multiple 
and highly--charged ions mainly because of the
relativistic contraction of the wave functions towards the nucleus. This
contraction can lower the cross sections easily by a factor of two or more
in the high--Z domain. The incorporation of higher multipoles beyond the 
E1--E1 dipole approximation, in contrast, contributes even for large 
values of $Z \,\sim\,100$ only to $\le\,$ 5\%{} for photon energies near 
the two--photon threshold. Somewhat larger retardation effects, however, 
are found for higher photon energies. For a two--photon excess energy of, say,
40~\%{} above the threshold [cf.\ Figure \ref{f:high_energy}],
the \textit{retarded} two--photon cross sections (solid line) are now 
larger than the cross sections in the long--wavelength approximation with 
deviation up to about 30~\%{} at the high--$Z$ end of the sequence. The 
behaviour of the retarded cross sections with respect to the long--wavelength 
approximation clearly shows the importance of higher multipoles which, 
otherwise, are usually seen only in angle--differential measurements 
(Surzhykov \etal{} 2002). 

\section{Conclusion}

\label{conclusion}

In conclusion, the non--resonant two--photon ionization of hydrogen--like ions 
has been studied in detail within the relativistic theory. Emphasize was placed,
in particular, on the relativistic contraction of the wave functions as well 
as on the retardation in the cross sections which arise from higher
multipoles of the radiation field. However, our computations also showed that 
a Green's function approach may provide a reliable access to second--order
properties other than the total two--photon ionization cross sections. 
Investigations on the angle--differential emission of electrons as well as 
the two--photon decay of few--electron ions are currently under work.

\section*{Acknowledgment:}

This work has been supported by the Deutsche Forschungsgemeinschaft (DFG) 
within the framework of the Schwerpunkt 'Wechselwirkung intensiver 
Laserfelder mit Materie'.

\section*{References:}

\begin{harvard}

\item[] Antoine P, Essarroukh N--E, Jureta J, Urbain X and Brouillard F 
        \,1996\, \JPB \textbf{29} 5367

\item[] Jaouen M, Laplanche G and Rachman A
        \,1984\, \JPB \textbf{17} 4643

\item[] Fritzsche S \,1997\, \textit{Comput. Phys. Commun.} \textbf{103} 51

\item[] Fritzsche S, Inghoft T, Bastug T and Tomaselli M \,2001\, 
\textit{Comput. Phys. Commun.} \textbf{139} 314

\item[] Karule E \,1985\, \JPB \textbf{18} 2207

\item[] Kornberg M~A, Godunov A~L, Ortiz S~I, Ederer D~L, McGuire J~H and
        Young L \,2002\, \textit{Journal of Synchrotron Radiation} \textbf{9} 
        298

\item[] Koval P and Fritzsche S \,2003\, \textit{Comput. Phys. Commun.} 
        in press

\item[] Laplanche G, Durrieu A, Flank Y, Jaouen M and Rachman A \,1976\,
        \JPB \textbf{9} 1263

\item[] Maquet A, Veniard V and Marian T~A \,1998\,
        \JPB \textbf{31} 3743

\item[] Rose M~E, \textit{Elementary Theory of Angular Momentum} \,1957\,
        (Wiley, New York)

\item[] Santos J~P, Patte P, Parente F and Indelicato P \,2001\, 
        \EJP D \textbf{13} 27.

\item[] Surzhykov A, Fritzsche S, Gumberidze A and St\"o{}hlker Th \,2002\, 
        \PRL \textbf{88} 153001

\item[] Swainson R~A and Drake G~W~F \,1991\, \JPA \textbf{24} 95

\item[] Szymanowski C, V\'e{}niard V, Ta\"{\i}{}eb R and Maquet A \,1997\,
        Europhys.~Lett. \textbf{6} 391

\item[] Zernik W \,1964\, \PR \textbf{135} A51

\end{harvard}

\end{document}